\documentclass[10pt,twocolumn, nofootinbib, notitlepage,aps,pra,showpacs,superscriptaddress,floatfix]{revtex4-1}

\usepackage{times, mathrsfs, amsmath, amsfonts, graphics, graphicx, cancel, color, amsthm, bbm, mathtools, amssymb, physics}
\usepackage[nice]{nicefrac}
\usepackage[english]{babel}
\usepackage[margin=.75in]{geometry}
\usepackage[T1]{fontenc} 
\usepackage{capt-of}

\usepackage{xfrac,relsize,float}
\usepackage{comment}

\usepackage{appendix}

\usepackage[unicode=true, breaklinks=false, pdfborder={0 0 1}, backref=false, colorlinks=true, linkcolor=blue, citecolor=blue]{hyperref}

\def\bra#1{\mathinner{\langle{#1}|}}
\def\ket#1{\mathinner{|{#1}\rangle}}

\def\BraVert{\egroup\,\mid\,\bgroup}

\definecolor{Blue}{rgb}{0,0,1}
\definecolor{Red}{rgb}{1,0,0}
\definecolor{Green}{rgb}{0,1,0}
\definecolor{darkgreen}{rgb}{0,.7,0}
\definecolor{Purp}{rgb}{.2,0,.2}
\definecolor{white}{rgb}{1,1,1}

\usepackage{bbold}

\begin{document}
\title{Probing quantum coherence at a distance and Aharonov-Bohm non-locality}

\author{Sebastian Horvat}
\affiliation{Faculty of Physics, University of Vienna, Boltzmanngasse 5, 1090 Vienna, Austria}

\author{Philippe Allard Gu\'{e}rin}
\affiliation{Faculty of Physics, University of Vienna, Boltzmanngasse 5, 1090 Vienna, Austria}
\affiliation{Institute for Quantum Optics and Quantum Information (IQOQI), Austrian Academy of Sciences, Boltzmanngasse 3, 1090 Vienna, Austria}
\affiliation{Perimeter Institute for Theoretical Physics, 31 Caroline St. N, Waterloo, Ontario, N2L 2Y5, Canada}

\author{Luca Apadula}
\affiliation{Faculty of Physics, University of Vienna, Boltzmanngasse 5, 1090 Vienna, Austria}
\affiliation{Institute for Quantum Optics and Quantum Information (IQOQI), Austrian Academy of Sciences, Boltzmanngasse 3, 1090 Vienna, Austria}

\author{Flavio Del Santo}
\affiliation{Faculty of Physics, University of Vienna, Boltzmanngasse 5, 1090 Vienna, Austria}
\affiliation{Institute for Quantum Optics and Quantum Information (IQOQI), Austrian Academy of Sciences, Boltzmanngasse 3, 1090 Vienna, Austria}
\affiliation{Basic Research Community for Physics (BRCP)}

\date{\today}
\date{\today}

\begin{abstract}
In a standard interferometry experiment, one measures the phase difference between two paths by recombining the two wave packets on a beam-splitter. However, it has been recognized that the phase can also be estimated via local measurements, by using an ancillary particle in a known superposition state. In this work, we further analyse these protocols for different types of particles (bosons or fermions, charged or uncharged), with a particular emphasis on the subtleties that arise when the phase is due to the coupling to an abelian gauge field. In that case, we show that the measurable quantities are spacetime loop integrals of the 4-vector potential, enclosed by two identical particles or by a particle-antiparticle pair. Furthermore, we generalize our considerations to scenarios involving an arbitrary number of parties performing local measurements on a general charged fermionic state. Finally, as a concrete application, we analyze a recent proposal involving the time-dependent Aharonov-Bohm effect ~\cite{marletto2020aharonov}.
\end{abstract}

\maketitle




\section{Introduction}\label{Introduction}
One ubiquitous feature that distinguishes quantum and classical theory is the superposition principle. 
In order to detect the presence of a quantum superposition (coherence), one has to perform measurements of at least two incompatible observables and, in fact, it is sufficient to measure the relative phase between the two (or more) states supposedly in superposition, in addition to the absolute value of the amplitudes. This can for example be achieved by an interferometric experiment, whose simplest implementation is the Mach-Zehnder interferometer. There, a single particle is prepared in superposition of two spatially separated paths (by sending it through a beam-splitter). The phase difference between the paths can be measured by recombining the paths at a second beam-splitter and by collecting the statistics of detectors placed at the output ports.

A celebrated variant of this interferometric experiment allows to detect a phase that is due to the interaction between a charged quantum particle (e.g. an electron) and the electromagnetic potential, known as the Aharonov-Bohm (A-B) effect~\cite{Aharonov1959, Aharonov1961, peshkin1989aharonov}. In particular, this experiment features a Mach-Zehnder interferometer that encircles a solenoid such that, despite the electromagnetic field being zero everywhere on the paths visited by the electron, there is a measurable phase that is directly proportional to the magnetic flux through the surface crossed by the solenoid. It thus seems that a relative quantum phase can only be measured indirectly by recombining the two paths in a closed interferometer, and arguably this limitation looks even more dramatic in the case of Aharonov-Bohm-like phases, since they depend on the value of the total electromagnetic flux contained in a closed region. Despite these indications, a number of works have proposed protocols to detect quantum superpositions without needing to reinterfere the beams, i.e. by using only local operations and classical communication (LOCC)~\cite{Aharonov2000, Vaidman2012, marletto2020aharonov, DelSanto2019, saldanha2019local, esin2019detection}.

In this work, we further analyze these kinds of protocols, characterizing their domain of applicability to different types of particles (bosons or fermions, charged or uncharged) and emphasize the constraints imposed by superselection rules and gauge symmetries in determining which observables are measurable (as discussed before in, e.g. Refs.~\cite{divincenzo2002quantum, kitaev2004superselection, van2006quantum, bartlett2007reference}). After a brief analysis of uncharged bosons, we focus on uncharged fermions for which the parity superselection rule forbids certain measurements. Nevertheless, one can circumvent this limitation by using an ancillary system as a resource; indeed, we show that an arbitrary number of parties sharing a generic fermionic pure state, can perform full state tomography by means of a known delocalized ancillary state and LOCC. 

We then proceed with the case of electromagnetically charged particles and explain issues arising from gauge coupling and gauge invariance. Contrarily to what the authors of Ref.~\cite{marletto2020aharonov} have recently proposed, we show that in protocols involving an ancillary system and local measurements, the measured phase corresponds always to the net phase picked up around a closed loop in spacetime. Thus, even if it is true that one can extract information about the surrounding electromagnetic field by means of only LOCC conducted at distant locations (i.e., without recombining the paths in an interferometer) and an ancillary system, its value is still equal to the electromagnetic flux through an hyper-surface enclosed by the paths travelled by the particle(s) in spacetime. We then generalize the latter result to an arbitrary number of sources and parties and show that all information about the gauge field that the parties can gather from LOCC can be reconstructed from bipartite loop integrals for all pairs of sources and parties.

In the final section, we explicitly apply our considerations to the scenario proposed in \cite{marletto2020aharonov, saldanha2019local}, which deals with the time-dependent Aharonov-Bohm effect. We thus show that, even if it is true that the Aharonov-Bohm phase is detectable by means of only LOCC conducted at distant locations (i.e., without recombining the paths in an interferometer), its measurable value is still equal to the (gauge-independent) electromagnetic flux through the hyper-surface enclosed by the paths traveled by the particle(s) in spacetime and as such it is not acquired locally.\footnote{The term ``non-locality'' has several meanings in quantum theory (see e.g.~\cite{Aharonov2000} for a concise overview). Indeed, it most customarily refers to the impossibility of explaining the statistical correlations of quantum measurements conducted at distant locations by means of local hidden-variables. Yet, in the context of the Aharonov-Bohm effect --and thus throughout this paper-- non-locality refers to a ``topological'' property of quantum theory. Namely, the fact that a measurable quantum phase in the presence of a gauge coupling is not acquired by summing up physically meaningful gauge-invariant quantities in a point-by-point (i.e. local) fashion along the path travelled by the particle. Rather, the phase due the gauge coupling can only be attributed to closed paths in spacetime, and it can be expressed as the net electromagnetic flux through a hypersurface enclosing that path.}

\section{Local measurements of the interferometric relative phase}
\label{localphases}
\begin{figure}[H]
\includegraphics[width=\linewidth]{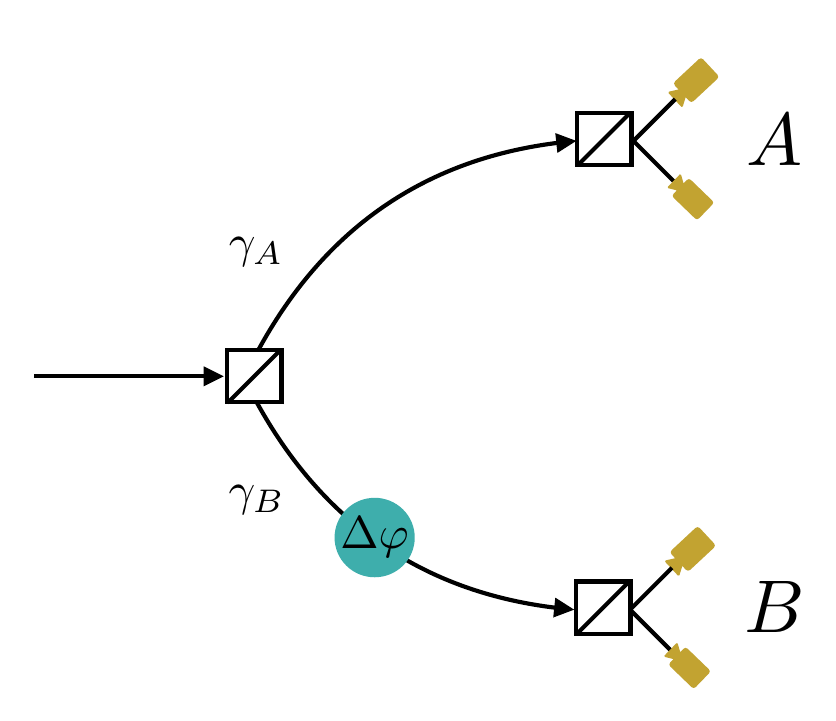}
\caption{A particle is sent through a 50/50 beam-splitter, along two paths, $\gamma_A$ and $\gamma_B$. For simplicity, we model the phase difference as arising due to a phase shifter placed along one of the paths. The goal of Alice and Bob is to measure the phase $\Delta \varphi$ using LOCC.}
\centering
\label{fig:mach_zehnder}
\end{figure}

In this section we review some protocols allowing to measure locally the relative phase acquired along two arms of an interferometer, without having to recombine the two paths. The setup we consider is illustrated in Figure~\ref{fig:mach_zehnder} and is similar to the settings studied in Refs.~\cite{marletto2020aharonov, DelSanto2019, saldanha2019local} .  Two experimenters, Alice and Bob, reside at two distant locations and a quantum particle is sent in an equal-weighted superposition of the two paths towards the parties. Most of the following applies to any particle type (boson or fermion, charged or uncharged); when specificities arise we will point them out explicitly.
We assume that the situation can be described  as a superposition over classical paths, $\gamma_A$ and $\gamma_B$, leading to Alice and Bob. In this case, the phase acquired along a path is simply related to the Lagrangian action $S$ of the corresponding classical trajectory. More precisely, let the phase difference be defined as (we are using units where $\hbar = 1$) 
\begin{equation}
\label{eq:delta_phi0}
\Delta \varphi = i (S(\gamma_B) - S(\gamma_A)) = i \int_{t_i}^{t_f} dt \big( \mathcal{L}(\gamma_B (t )) - \mathcal{L}(\gamma_A(t)) \big),
\end{equation}
where $t_i$ is the time when the particle is emitted from the beam-splitter and $t_f$ is the time when the parties receive the particle. The shared state between the two parties at the final time $t_f$ is then
\begin{equation}
\ket{\psi}=\frac{1}{\sqrt{2}}\left(\ket{A}+e^{i\Delta \varphi }\ket{B}\right),
\end{equation}
where $|A\rangle$ and $|B\rangle$ correspond to states where the particle is localised at Alice's or Bob's position, respectively. When dealing with interferometric experiments it is convenient to adopt a second-quantized notation:
\begin{equation}
\ket{\psi}=\frac{1}{\sqrt{2}}\left(a^{\dagger}_{A}+e^{i\Delta\varphi}a^{\dagger}_{B}\right)\ket{0},
\end{equation}
where $a^{\dagger}_{A/B}$ represent creation operators (either bosonic or fermionic), creating one particle at Alice's and Bob' location respectively, and $\ket{0}$ is the vacuum state. 

The task consists in estimating the phase difference $\Delta \varphi$, using only LOCC. This would be straightforward if Alice and Bob were able to implement local projective measurements that involve the preparation of superposition of different number states, i.e., of the form $\Pi_{\pm}=\ket{\pm}\bra{\pm}$, where
\begin{equation}\label{boson_measurement}
\ket{\pm}_{A/B}=\frac{1}{\sqrt{2}}\left( \mathbb{1} \pm  a^{\dagger}_{A/B}  \right)\ket{0}.
\end{equation}
This, however, is only possible for uncharged bosons, because otherwise this measurement is forbidden by the parity and/or charge superselection rules~\cite{Wick1952, Wick1970, Strocchi1974, Wightman1995}. For photons, these measurements can be carried out by making each local mode interact with another system such as an atom, and then locally measuring the two atoms in an appropriate basis \cite{VanEnk2005, Aharonov2000, Fuwa2015, Rosenfeld2017}, or by measuring local interference with a coherent state \cite{Aharonov2000}. However, for general species of particles, the measurement of Eq.~\eqref{boson_measurement} is fundamentally forbidden (by specific superselection rules) and the determination of the relative phase is apparently impossible.

\subsection{A general method for local phase measurements}
\label{sec:gen_protocol}

Despite the above-mentioned issues, a possible solution to the apparent impossibility of locally measuring phase differences of fermions has been proposed in Refs.~\cite{Aharonov2000, marletto2020aharonov}. The idea is to bypass the superselection rule by using an ancillary particle in a known superposition state as a resource which enables the phase measurement. The protocol proceeds as follows. Let us assume that Alice and Bob already possess an ancillary particle in spatial superposition of their locations and that the phase difference between the two components of the wave function is null.\footnote{As we will discuss in detail in Section~\ref{sec:charged}, for charged particles this assumption is not as innocent as it may look, for it corresponds to fixing a gauge.} 
After some time, the second particle is sent in spatial superpostition of the two paths, of which we want to measure the relative phase difference.
Upon reaching Alice and Bob, the total state of the two-particle system is
\begin{equation}
\label{eq:ancilla}
\ket{\psi}=\frac{1}{2}\left(a^{\dagger}_{A_1}+ a^{\dagger}_{B_1}\right)\left(a^{\dagger}_{A_2}+e^{i \Delta\varphi} a^{\dagger}_{B_2}\right)\ket{0},
\end{equation}
where the mode corresponding to the ancillary particle is indicated by subscript 1 and the mode corresponding to the second one --whose relative phase needs to be estimated-- by subscript 2. Both Alice and Bob now possess two ``wave packets'', one arising from each particle. Furthermore, suppose they possess local beam-splitters allowing them to locally interfere their wave packets, and detectors allowing to measure the outcome statistics at the output ports of the beam-splitters. In order for the parties to be able to perform the measurements, the particles must have the same parity and charge because of the associated superselection rules. We thus choose the particles to be identical. Now, suppose that Alice and Bob perform their measurements and discard the results if they detect either zero or two particles: the postselected quantum state of interest is then (prior to the interference through the beam-splitter)
\begin{equation}
\ket{\psi}_{PS}=\frac{1}{\sqrt{2}}\left( a^{\dagger}_{B_1}a^{\dagger}_{A_2}+e^{i\Delta\varphi}a^{\dagger}_{A_1}a^{\dagger}_{B_2} \right)\ket{0},
\end{equation}
The beam-splitters, followed by measurements at the output ports, implement local projective measurements of the form $\Pi_{\pm}=\ket{\pm}\bra{\pm}$ acting on $\ket{\psi}_{PS}$, where
\begin{equation}\label{fermion_measurement}
\ket{\pm}_{A/B}=\frac{1}{\sqrt{2}}\left( a^{\dagger}_{A_1/B_1} \pm  a^{\dagger}_{A_2/B_2}  \right)\ket{0}.
\end{equation}
Note that the outcome probabilities  do not depend on whether the particles are fermions or bosons, because we are only concerned with the subset of events in which a single particle is found at each of the two locations. The outcome statistics enables Alice and Bob to reconstruct the phase shifter's phase $\Delta \varphi$ using only LOCC despite issues caused by the parity superselection rule.

The latter considerations were concerned with the case of two parties sharing one particle. In Appendix~ \ref{app:A}, we provide a generalization to a scenario involving $N$ parties sharing a general fermionic uncharged state, which can now involve an arbitrary number of excitations. Notice that the parity superselection rule allows the state to be in a superposition of states with different numbers of fermions (albeit with equal parity). We show that, analogously to the single-particle case, the parties can fully reconstruct an arbitrary fermionic pure state using LOCC and a common global state as a resource. The protocol essentially involves the usage of an auxiliary reference state prepared by an external party and local beam-splitter operations performed by the $N$ parties. Even though the parties' measurements are local, the overall process involves the preparation of a global (entangled) state which cannot be generated by local means; this is in accord with the results of Refs.~\cite{d2014fermionic, Lugli2020}.


\section{Charged particles and the gauge-dependence of relative phases} \label{sec:charged}

Let us now turn to charged particles, for which we already pointed out that measurements of the type in expression \eqref{boson_measurement} are prohibited by the charge superselection rule. Furthermore, in the general protocol of Section~\ref{sec:gen_protocol}, we had to assume, in Eq.~\eqref{eq:ancilla}, that it is possible to prepare the ancillary particle in a known state. As we will emphasize in this section, since the phase acquired along a path is gauge-dependent, the condition of Eq.~\eqref{eq:ancilla}, namely that the ancillary state has zero relative phase, is not a gauge-independent property. In the following we show how the protocol can nevertheless be used to provide local measurements of gauge-invariant phases, i.e. phases that are acquired by integrating along a closed loop in spacetime. Hereinafter, we will focus on charged fermions, and, as a matter of simplicity, we will consider the particular case of electrons. 
Consider once again the setup of Fig.~\ref{fig:mach_zehnder}. The phase difference is given by Eq.~\eqref{eq:delta_phi0}, with Lagrangian
\begin{equation} \label{eq:L}
\mathcal{L}=\frac{1}{2m} \vec{\dot{x}}^2 - e \vec{\dot{x}} \cdot \vec{A} + eV\left(\vec{x}\right). 
\end{equation}
We emphasize that setting the gauge potential to zero, even in the absence of external electric and magnetic fields, amounts to fixing a gauge. Hence, the accumulated phase on the two paths now necessarily depends also on the electromagnetic potential:
\begin{equation}
\ket{\psi}=\frac{1}{\sqrt{2}}\left(e^{i e\int_{\gamma_A} A^{\mu}dx_{\mu}}\ket{A}+e^{i e\int_{\gamma_B} A^{\mu}dx_{\mu}}e^{i\beta}\ket{B}\right),
\end{equation}
where $\beta$ is the gauge-invariant "mechanical" phase difference, i.e. due to the kinetic term $\frac{1}{2m} \vec{\dot{x}}^2$ in the Lagrangian; whereas $A^\mu$ is the 4-vector potential, i.e. (in the units where $c=1$) $A^\mu = (V, \vec{A})$, and we use a metric with signature $(+, -,-,-)$. Therefore, the phase difference between the two paths now reads
\begin{equation}
\Delta\varphi=\beta+e\int_{(\gamma_B-\gamma_A)} A^{\mu}dx_{\mu},
\end{equation} 
which is an explicitly gauge-dependent quantity. We thus see why it is necessary to introduce the charge superselection rule which forbids us to implement the projectors of Eq.~\eqref{boson_measurement}: if this rule did not exist we would be able to measure physically meaningless gauge-dependent quantities!\footnote{A similar argument has already been invoked for example in \cite{erez2010ab}; for a more formal treatment of the relationship between the charge superselection rules and gauge invariance, see \cite{earman}.} This argument has two main consequences: (i) the principle of gauge invariance implies the charge superselection rule and prohibits this type of local tomographic protocols for a delocalized electron. Furthermore, (ii) the acquired phase difference between the two paths is not an observable since it is a gauge-dependent quantity: it might be equal to the mechanical phase shift $\beta$ only in a specific gauge.

On the other hand, note that if the electron's wave packets are reinterfered on a beam-splitter as it happens in a standard Aharonov-Bohm experiment, the phase difference between the two paths connecting the two beam-splitters reduces to 
\begin{equation}
\Delta\varphi=\beta+e\oint A^{\mu}dx_{\mu},
\end{equation}   
where the loop integral is performed around the whole interferometer. The phase is now a gauge-independent quantity and it depends on the distribution of electromagnetic currents in spacetime. A specific case is a regular Mach-Zehnder interferometer, where, since there are no currents, the loop integral vanishes and, as expected, the total phase difference is equal to the mechanical phase shift $\beta$ only.
\subsection{Measuring phase differences at a distance}\label{Distance}
\begin{figure}[t]
\includegraphics[width=8cm]{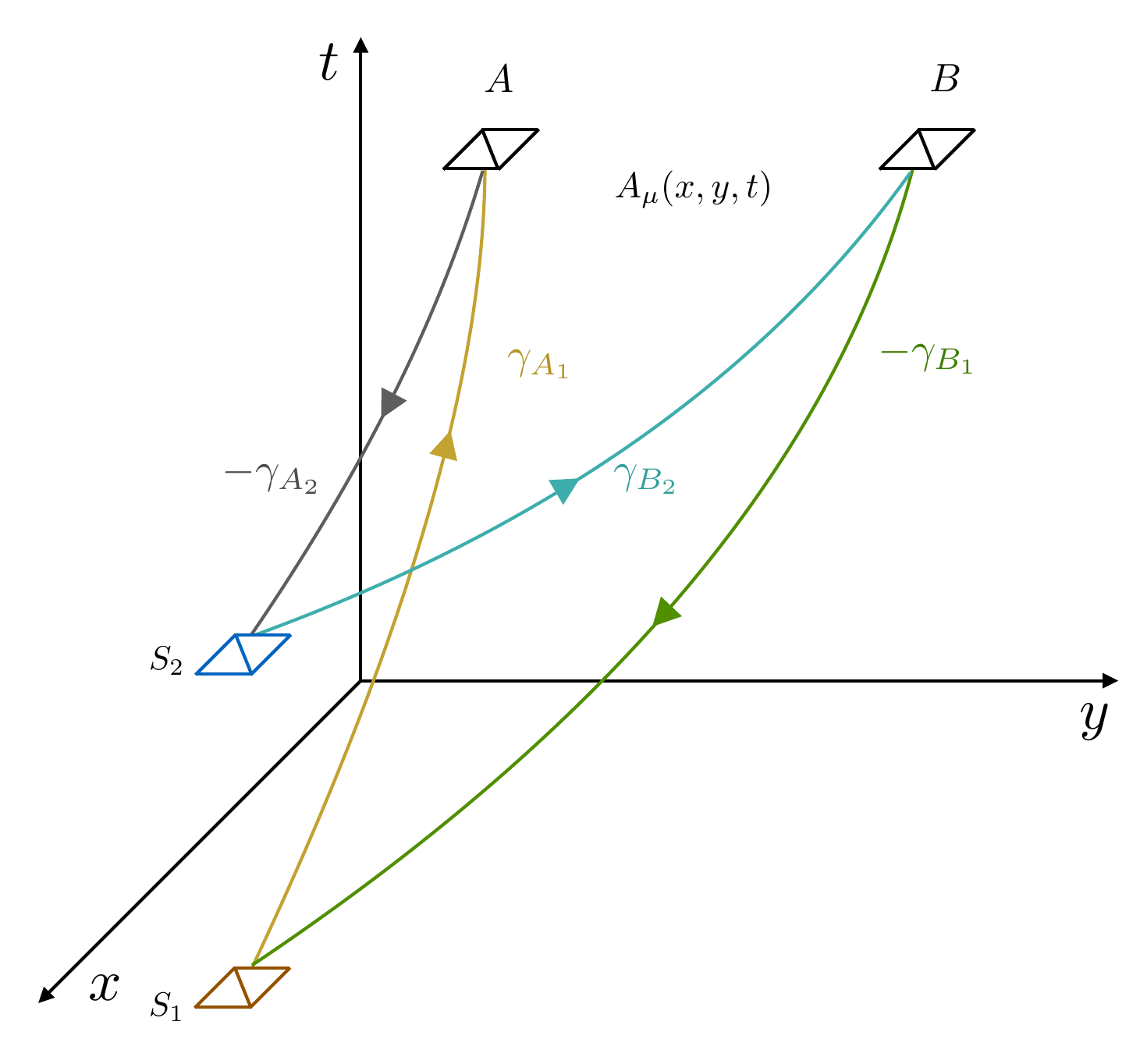}
\caption{Spacetime diagram of the protocol described in Section~\ref{Distance}. Two  identical particles are prepared at locations $S_1$ (resp. $S_2$) in an equal superposition of paths $\gamma_{A_1}$ and $\gamma_{B_1}$ (resp. $\gamma_{A_2}$, $\gamma_{B_2}$). The parties perform local measurements which allow them to reconstruct the phase $\Delta \varphi$ of Eq.~\eqref{eq:phase}, which is a gauge-invariant spacetime loop integral around the shown path (the arrows are drawn in the direction of integration of the loop integral).}
\centering
\label{fig:spacetime_AB}
\end{figure}

We now show how, as in the protocol of Section~\ref{sec:gen_protocol}, we can exploit an auxiliary identical particle (another electron) in order to circumvent the limitation imposed by the charge superselection rule and gauge invariance. The crucial difference with respect to the previous case is that, as already emphasized, there is an inevitable coupling to the gauge potential regardless of the surrounding electromagnetic sources, which implies that even the ancillary particle necessarily acquires a gauge-dependent phase difference which may be null only in a specific gauge (i.e. we cannot assume that Alice and Bob  can prepare the ancillary particle in a known state without fixing a gauge). After the particles are sent to the parties, the two-particle state is
\begin{equation}
\begin{split}\label{glbstate}
\ket{\psi}=\frac{1}{2}\left(e^{i e\int_{\gamma_{A_1}} A^{\mu}dx_{\mu}}a^{\dagger}_{A_1}+e^{i e\int_{\gamma_{B_1}} A^{\mu}dx_{\mu}}e^{i\beta_1}a^{\dagger}_{B_1}\right)\\
\left(e^{i e\int_{\gamma_{A_2}} A^{\mu}dx_{\mu}}a^{\dagger}_{A_2}+e^{i e\int_{\gamma_{B_2}} A^{\mu}dx_{\mu}}e^{i\beta_2}a^{\dagger}_{B_2}\right)\ket 0,
\end{split}
\end{equation}
where the mode corresponding to the first particle is indicated by subscript 1 and the second one by subscript 2, the paths $\gamma_{A_{1}}, \gamma_{A_{2}}, \gamma_{B_{1}}$ and $\gamma_{B_{2}}$ are defined in Fig. \ref{fig:spacetime_AB}, and $\beta_i$ is the mechanical phase difference between paths $\gamma_{B_i}$ and $\gamma_{A_i}$. Alice and Bob again perform measurements with local beam-splitters and postselect on the one-particle subsector:
\begin{equation}
\ket{\psi}_{PS}=\frac{1}{\sqrt{2}}\left( a^{\dagger}_{A_2}a^{\dagger}_{B_1}+e^{i\Delta\varphi}a^{\dagger}_{A_1}a^{\dagger}_{B_2} \right)\ket{0},
\end{equation}
where the cumulative phase difference is now given by
\begin{equation}\label{eq:phase}
\Delta\varphi=\beta_2 - \beta_1 +\varphi_{A_1}+\varphi_{B_2}-\varphi_{A_2}-\varphi_{B_1}, 
\end{equation}
with 
\begin{equation}\label{eq:phase2}
\varphi_{A_{i}}\equiv e\int_{\gamma_{A_{i}}} A^{\mu}dx_{\mu},
\end{equation}
where $i\in\{ 1, 2\}$ labels the paths on Alice's side, and with an analogous expression for $\varphi_{B_{i}}$. As in Section \ref{sec:gen_protocol}, the quantity $\Delta\varphi$ can be inferred from the statistics of the detection clicks at the two beam-splitters, respectively situated at Alice's and Bob's locations.

In Figure~\ref{fig:spacetime_AB} we portray the paths of the two particles in a spacetime diagram and show that they enclose a closed path. Therefore, the additional phase accumulated because of the interaction with the electromagnetic potential is gauge invariant. If there are no surrounding currents or fields, the loop integral vanishes and only the mechanical phase shift remains. On the contrary, if the particles are surrounded by an arbitrary distribution of currents, the loop integral does not generally vanish and can depend explicitly on the trajectory travelled by the ancillary particle (even if it is not acted on by classical forces).

Had the two particles different charges $e_1$ and $e_2$, the total phase $\Delta \varphi$ would involve the sum of two paths which do not compose into a spacetime loop integral as in \eqref{eq:phase} and would thus not be a gauge invariant quantity. However, in that case, the required measurement would not be possible since non-identical particles do not interfere with each other. Therefore, we see a consistency between gauge invariance, superselection rules and the operational attainability of the required local measurements.

In Appendix~\ref{app:positron}, we analyse a variation of the above experiment introduced by Aharonov and Vaidman in \cite{Aharonov2000}, which instead of two identical particles (e.g. two electrons) involves a particle-antiparticle pair (e.g. an electron and a positron). We show that the same quantity $\Delta \varphi$ from Eq. \ref{eq:phase} can be estimated by measuring photons resulting from the annihilation of the electron with the positron.\\
From these considerations we draw the following conclusions: 
\begin{itemize}
\item loops  (in spacetime) can be closed by states involving two identical particles or by a particle-antiparticle pair, i.e. by two different excitations of the same quantum field (contrary to the standard Aharonov-Bohm effect where the loop is closed by a single electron);
\item the paths traced by the two particles can be ``glued'' together via local (interference) measurements, or via particle-antiparticle annihilation;
\item in the presence of gauge coupling, the distinction between ``primary'' particle (whose phase we are trying to estimate) and ``ancillary'' particle (which serves to circumvent the superselection rule) is not well defined and can be manifested only in a specific gauge; the measured quantity is a collective property of the two excitations, namely the spacetime loop integral.
\end{itemize}

\subsection{General case}\label{general case charged} 
In Section \ref{sec:gen_protocol} and in Appendix \ref{app:A} we saw the possibility of performing state tomography of an arbitrary uncharged fermionic state using LOCC and an ancillary system. Now, we want to analyse a similar scenario in the case of charged particles. However, since the concepts of ``primary'' and ``ancillary'' systems are not well defined in the presence of gauge coupling, here we ask a different question. Suppose that an arbitrary number of parties perform local measurements on an arbitrary state of multiple charged particles: what are the spacetime loop integrals that the outcome probabilities depend on? Can all probabilities be reconstructed from loops similar to the one depicted in Fig. \ref{fig:spacetime_AB}? In Appendix \ref{app:C} we show that this is indeed the case. More precisely, suppose that we have $d$ sources emitting single identical charged fermions at $d$ spacetime points. Each source prepares a single particle in an arbitrary superposition of spatial trajectories. Moreover, suppose that there are $N$ parties who perform local number-preserving (linear) operations and measurements at $N$ space-like separated points located on a hypersurface which lies in the future of the $d$ sources. We show that the joint probability of the local measurement outcomes can be fully reconstructed from loop integrals as in Fig. \ref{fig:spacetime_AB} for all pairs of sources and all pairs of parties. This result shows that all information about the gauge field that is acquirable via local measurements on single particle excitations can be reduced to simple experiments involving two parties and two sources as in Fig. \ref{fig:spacetime_AB}.

\section{Non-local generation of the Aharonov-Bohm phase}\label{AB}


The traditional interpretation of the Aharonov-Bohm effect~\cite{Aharonov1959, Aharonov1961, peshkin1989aharonov} is that the electric and magnetic fields are in general not sufficient for describing the physics of certain (quantum) scenarios and that the gauge-theoretic electromagnetic potential is in fact indispensable. Going against the received view, Vaidman has argued that the Aharonov-Bohm phase can be explained without the introduction of potentials if one takes into account the quantum nature of the solenoid~\cite{Vaidman2012}. A weakness in his treatment is that it relies on an instantaneous interaction between the solenoid and the electron. Marletto and Vedral~\cite{marletto2020aharonov} --followed by further developments by Saldanha \cite{saldanha2019local}-- have recently addressed this problem, concluding that the Aharonov-Bohm phase is acquired locally. In this section we follow up on these recent developments by applying the analysis from the previous section to the specific case where the only source of electromagnetic fields is a solenoid with a time-dependent current. We show that that the Aharonov-Bohm phase is not acquired locally, in the sense that the only measureable quantities involved in this type of experiments correspond to the integral of the 4-vector potential around whole \textit{spacetime loops}. The phase acquired along smaller portions of the particle's path is not measureable.

\begin{figure}[t]
\includegraphics[width=8cm]{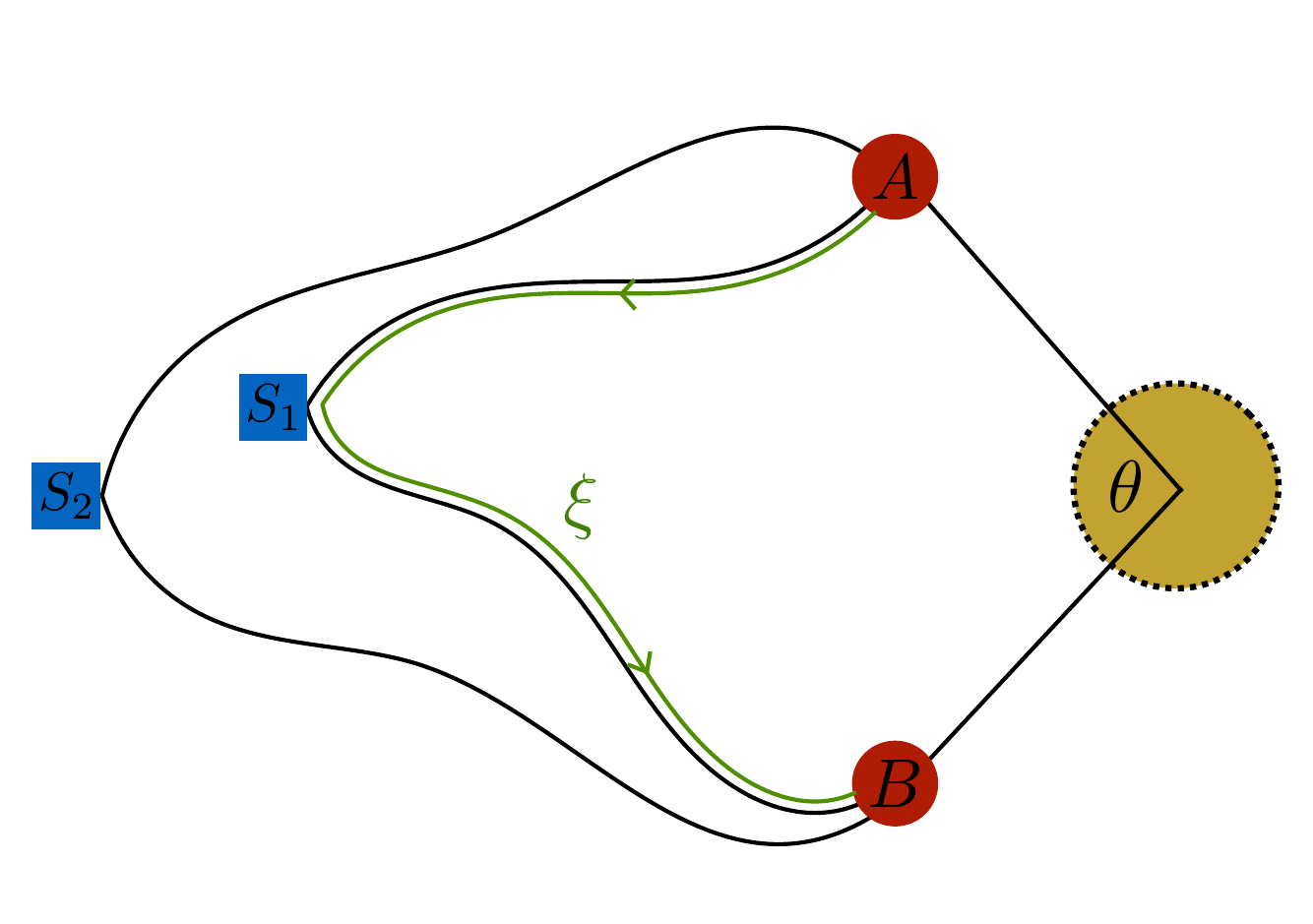}
\caption{Spatial representation of the scenario of Fig.~\ref{fig:spacetime_AB}, for the special case studied in Ref.~\cite{marletto2020aharonov}, where the source of $A^\mu$ is a solenoid with a time-dependent current. At time $t=0$ the flux through the solenoid is zero, and a particle is prepared at $S_1$ in an equal superposition of paths $\gamma_{A_1}$ and $\gamma_{B_1}$. Later, this particle is trapped at locations $A$ and $B$, following which the flux is increased up to $\Phi_f$. Finally, the second particle is sent at $S_2$ and each party performs a local measurement on their ``wave packets''.}
\centering
\label{fig:AB}
\end{figure}

The scenario studied in Ref.~\cite{marletto2020aharonov} is a special case of the general situation in Fig.\ref{fig:spacetime_AB}. During the journey of the ancillary particle towards the parties, the flux through the coil is set to zero. After the two wave packets of the ancillary particle arrive at the location of the parties, they are trapped using some external field. The current (and thus the flux) in the coil is then increased slowly until it reaches a stationary state, following which the second particle is sent to the parties and the local measurements of~\eqref{fermion_measurement} are performed. The detection probabilities depend on $\Delta \varphi$ as defined in~\eqref{eq:phase}, where (in the Coulomb gauge) the vector potential in the region outside the coil is given by
\begin{equation}\label{A}
\vec{A}(\vec r, t) =\frac{\Phi(t)}{2\pi (x^2+y^2)}\vec r \times \hat{z},
\end{equation}
where $\Phi(t)$ is the magnetic flux through the solenoid at time $t$, and $\vec r$ is the position vector of the particle (with the origin of the coordinates centered in the solenoid). Note that the electric field is not zero outside the solenoid at all times, because of the time-dependence of the current in the solenoid. Since the charge density is zero everywhere and at all time, we have that the scalar potential is zero in the Coulomb gauge, and thus the phase that the ancillary particle accumulates due to the solenoid during the time when it is trapped in Fig.~\ref{fig:spacetime_AB} is zero (in this gauge). Furthermore, when the ancillary electron is moving, there is no current in the coil and consequently no 4-vector potential is interacting with the charge, so the contribution of the solenoid to the phase accumulated along $\gamma_{A_{1}/B_{1}}$ is zero. Hence, we are left with the phase acquired by the second particle
\begin{equation}
\label{eq:phase_xi}
\Delta \varphi =e\int_{\xi}\vec{A} \cdot d \vec{l},
\end{equation}
where $\xi$ is the path shown in Figure~\ref{fig:AB}. Using the expression~\eqref{A} for the vector potential and switching to polar coordinates yields
\begin{equation}
\Delta \varphi =\int_0^\theta\frac{\Phi_f}{2\pi}d\theta'=\frac{\Phi_f}{2\pi}\theta,
\end{equation}
where $\Phi_f$ is the final flux through the solenoid and $\theta$ is the angle defined in Figure~\ref{fig:AB}. The probabilities for the measurement described in Section~\ref{Distance} depend on $\Delta \varphi$ and thus the outcome statistics allows to estimate $\Delta \varphi$ using LOCC.

It is tempting to interpret the calculation that we have just done, in particular Eq.~\eqref{eq:phase_xi}, as showing that the phase is locally accumulated along the path $\xi$. However, this apparent localization of the phase accumulation is merely a consequence of our (arbitrary) choice to work in the Coulomb gauge. In other gauges, the phase accumulated by the ancillary particle is not necessarily zero: only the full loop integral is gauge-independent. Notice that, even though the particles seemingly enclose a spatial region with no electromagnetic fields, the measured quantity $\Delta \varphi$ corresponds to the spacetime loop integral which is non-zero due to the induced electric field (which arises due to the time dependence of the current) piercing through the hypersurface enclosed by the spacetime loop.

\section{Conclusions and Outlook}

We have studied a general protocol allowing to locally measure the relative phases of the state of a particle that is prepared, by using a beam-splitter, in a superposition of different spatial locations. In order to perform this protocol, the localised parties must share a known superposition state of an ancillary particle. We have shown how, using this known state as a resource, one can estimate the relative phase by performing only local measurements, despite the restrictions imposed by the parity superselection rule. Moreover, we extended the protocol to general fermionic systems shared by an arbitrary number of parties and we showed how the parties can perform full state tomography using LOCC and a global state as a resource.
We proceeded by addressing the case of electromagnetically charged particles, where the unavoidable coupling to the gauge field makes it impossible to control the relative phase of the ancillary state without fixing the gauge. We have shown that the protocol nevertheless measures a gauge-independent quantity, namely a spacetime loop integral of the 4-vector potential. The protocol requires the two particles to possess equal charges in order to obtain gauge invariant quantities; however, the required measurements would be physically impossible on two different charges due to the charge superselection rule. Alternatively, one can employ a slightly modified protocol that involves particles with opposite charges (e.g. electron and positron) which can annihilate into a pair of uncharged particles (photons): the obtained measurement results yield the same gauge invariant phase as in the former protocol. The latter discussion shows the tight relation between charge conservation, gauge invariance and superselection rules. We then proceeded with the general scenario involving many source and many parties performing LOCC and showed that all probabilities arising from such experiments can be reduced to simple combinations of bipartite loop integrals involving two sources and two parties. Finally, we have applied the protocol to the case of local measurements of phases in the time-dependent Aharonov-Bohm effect, and, in particular, we have demonstrated its application to the setups of Refs.~\cite{marletto2020aharonov, saldanha2019local}. Since all probabilities obtained in these experiments depend on loop integrals which are explicitly non-local quantities, the interpretation that the phase is acquired locally is not viable, and is apparently manifested only in a specific gauge.

In this manuscript we focused on scenarios which involve quantum particles coupled to Abelian gauge fields; it would be interesting to see to what extent our analysis can be extended to the non-Abelian case, where the probabilities would depend on gauge-invariant functionals of Wilson loop operators. In particular, one should inspect whether the result of section \ref{general case charged} holds in any gauge theory or only in the Abelian ones.


\begin{acknowledgements}
We would like to thank \v{C}aslav Brukner and Borivoje Daki\'c for useful discussions and comments that greatly improved this paper. 
 F.D.S. acknowledges the financial support through a DOC Fellowship of the Austrian Academy of Sciences (\"OAW). S.H acknowledges support from an ESQ Discovery Grant of the Austrian Academy of Sciences (\"OAW). P.A.G. and L.A. acknowledge support from the research platform Testing Quantum and Gravity Interface with Single Photons (TURIS), the Austrian Science Fund (FWF) through the projects BeyondC (F7113-N48), and  the doctoral  program Complex Quantum Systems (CoQuS) under Project No. W1210-N25 and the financial support from the EU Collaborative Project TEQ (Grant Agreement No.766900). They also acknowledge a grant from the Foundational Questions Institute (FQXi) Fund and by the ID\# 61466 grant from the John Templeton Foundation, as part of the "The Quantum Information Structure  of  Spacetime  (QISS)"  Project  (qiss.fr). The opinions expressed in this publication are those of the author(s) and do not necessarily reflect the views of the John Templeton Foundation.
\end{acknowledgements}

\bibliography{biblio}


\appendix

\begin{widetext}

\section{Local measurements on generic fermionic uncharged systems}\label{app:A}

Let us start by considering a simple generalisation of Fig.~\ref{fig:mach_zehnder} to $N$ parties. The particle goes through a beam-splitter with $N$ output ports, and along each arm it acquires an unknown phase $\varphi_i$, so that the state received by the observers is $ \frac{1}{\sqrt{N}} \sum_{i = 1}^N e^{i \varphi_i} c^\dagger_i |0 \rangle$, where $c^\dagger_i$ is an operator that creates a particle at the location of observer $i$. We would like to perform a procedure for estimating the phases $\varphi_i$ that uses only local operations and classical communication. This can be achieved if the parties share a second particle in a known state $\frac{1}{\sqrt{N}} \sum_{i = 1}^N c^\dagger_i |0 \rangle$. Let each party apply a beam-splitter locally and measure at the click at the output ports, and post-select on cases where the two particles are found at different locations; this happens with probability $1 - \frac{1}{N}$. Supposing that the particles are found at positions $i$ and $j$, the probabilities for each of the 4 possible outcomes are the same as in the case of Fig.~\ref{fig:mach_zehnder}, so the above protocol allows to give an estimate for $\varphi_i - \varphi_j$. After performing many rounds it will be possible to reconstruct all the phases $\varphi_k$ with a good accuracy. An interesting feature of this protocol is that the post-selection probability goes to one as the number of parties becomes large, which means that in this limit almost all rounds of the experiment yield useful information (in contrast with the two arm case, where half of the rounds have to be discarded).

Turning now to the general case, the most general fermionic state shared by $N$ parties is
\begin{equation}\label{inital state0}
\ket{\psi}=\left\{\sum_{\vec{x}_1,...,\vec{x}_N} \lambda\left(\vec{x}_1,...,\vec{x}_N\right) e^{i\varphi\left(\vec{x}_1,...,\vec{x}_N \right)} \prod_{j_1} \left(c_{j_1}^{(1)\dagger}\right)^{x_{1j_1}} \prod_{j_2} \left(c_{j_2}^{(2)\dagger}\right)^{x_{2j_2}}...\prod_{j_N} \left(c_{j_N}^{(N)\dagger}\right)^{x_{Nj_N}} \right\}\ket{0}.
\end{equation}
In the latter expression the sum ranges over all bit strings, the length of which depends on the maximal number of local modes available to each of the parties (the bit string notation automatically implements the fact that there can be no more than one excitation per mode). $\lambda\left(\vec{x}_1,...,\vec{x}_N\right)$ are real amplitudes, $\varphi\left(\vec{x}_1,...,\vec{x}_N \right)$ are the mechanical phases that we want to estimate, and $c_{j_k}^{(k)\dagger}$ denote fermionic creation operators that create one fermion in the $j_k$-th mode of the $k$-th party.\\
In order to measure the weights $\lambda\left(\vec{x}_1,...,\vec{x}_N\right)$, each party performs a local projective measurement on their local modes, yielding the desired information via 
\begin{equation}
\lambda (\vec{x}_1,...,\vec{x}_N)=|\bra{\psi} \prod_{j_1} \left(c_{j_1}^{(1)\dagger}\right)^{x_{1j_1}}...\prod_{j_N} \left(c_{j_N}^{(N)\dagger}\right)^{x_{Nj_N}}\ket{0}|.
\end{equation}
The parties communicate their local results to an external party who estimates the amplitudes $\lambda\left(\vec{x}_1,...,\vec{x}_N\right)$, prepares a uniform superposition over states with non-zero amplitude and sends it towards the $N$ parties, who store it in separate modes from the ones occupied by the original state. The "copied" state is thus
\begin{equation}
\ket{{\tilde{\psi}}}=\frac{1}{\sqrt{M}}\left\{\sum_{\vec{x}_1,...,\vec{x}_N}\prod_{j_1} \left(\tilde{c}_{j_1}^{(1)\dagger}\right)^{x_{1j_1}}...\prod_{j_N} \left(\tilde{c}_{j_N}^{(N)\dagger}\right)^{x_{Nj_N}} \right\}\ket{0},
\end{equation}
where $\tilde{c}_{j_k}^{(k)\dagger}\equiv c_{n_k+j_k}^{(k)\dagger}$, with $n_k$ being the maximum number of modes present at the $k$-th party's location and $M$ is the total number of non-zero components in the original state \eqref{inital state0}.\\
Next, each party interferes each of the original modes with the corresponding copied modes on local beam-splitters, i.e. $\forall k=1,...,N$ and $\forall j_k=1,...,n_k$
\begin{equation}
\begin{split}
c_{j_k}^{(k)\dagger} \rightarrow \frac{1}{\sqrt{2}} \left(c_{j_k}^{(k)\dagger}+\tilde{c}_{j_k}^{(k)\dagger} \right),\\
\tilde{c}_{j_k}^{(k)\dagger} \rightarrow \frac{1}{\sqrt{2}} \left(c_{j_k}^{(k)\dagger}-\tilde{c}_{j_k}^{(k)\dagger} \right).
\end{split}
\end{equation}
The final joint state, after undergoing the beam-splitter operations, is thus
\begin{equation}
\ket{\psi'}=\left\{\frac{1}{\sqrt{M}}\sum_{\substack{\vec{x}_1,...,\vec{x}_N \\\vec{x}_1',...,\vec{x}_N'}}2^{-\frac{1}{2}\left[\sum_{nj}\left(x_{nj}+x_{nj}'\right)\right]} \lambda_xe^{i\varphi_x} \prod_{k=1}^{N}\left[\prod_{j_k=1}^{n_k}\left( c_{j_k}^{(k)\dagger}+\tilde{c}_{j_k}^{(k)\dagger}\right)^{x_{kj_k}}\right]\prod_{k'=1}^{N}\left[\prod_{j_k'=1}^{n_k'}\left( c_{j_k'}^{(k')\dagger}-\tilde{c}_{j_k'}^{(k')\dagger}\right)^{x_{k'j_k'}'}\right]\right\}\ket{0},
\end{equation}
where $\lambda_x$ and $\varphi_x$ stand for $\lambda\left(\vec{x}_1,...,\vec{x}_N\right)$ and $\varphi\left(\vec{x}_1,...,\vec{x}_N\right)$.\\
Finally, the parties perform local projective measurements of their modes in the occupation number basis; the outcome probabilities depend explicitly on the required phases:
\begin{equation}
P_{xy}=\frac{1}{M2^{\left[\sum_{n,m}\left(x_{nm}+y_{nm}\right)\right]}}|\lambda_{x}+(-1)^{s_{xy}}e^{i\left(\varphi_{y}-\varphi_{x}\right)} \lambda_{y}|^2,
\end{equation}
where $x=(\vec{x}_1,...,\vec{x}_N)$ and $y=(\vec{y}_1,...,\vec{y}_N)$ are two different bit strings, $P_{xy}$ is the probability of projecting on state $\left\{ \prod_{k=1}^{N}\left[\prod_{j_k=1}^{n_k}\left(c_{j_k}^{(k)\dagger}\right)^{x_{kj_k}}\right]\prod_{k'=1}^{N}\left[\prod_{j_k'=1}^{n_k'}\left(\tilde{c}_{j_k'}^{(k')\dagger}\right)^{y_{k'j_k'}}\right]\right\}\ket{0}$, and $s_{xy}$ is an integer that arises due to fermionic anticommutation relations. The parties can thus estimate all the unknown phases $\varphi\left(\vec{x}_1,...,\vec{x}_N \right)$ using LOCC and a shared global ancillary state.

\section{The protocol involving a particle-antiparticle pair}
\label{app:positron}
Instead of two identical particles, one can choose the two particles used in the bipartite protocol of Section \ref{sec:charged} to be a particle-antiparticle pair, for instance, an electron and a positron (this possibility was discussed already in Ref. \cite{Aharonov2000}). In this case, Alice and Bob can annihilate their wave packets into photon pairs and measure the phase difference between the two components by performing local tomography on the resulting photons in a similar fashion as in section \ref{Distance}. More precisely, the state of the two particles before the annihilation process is the same as in protocol involving two identical particles, up to a minus sign in the phases acquired by the antiparticle (the ``ancillary'' system is now a positron with charge $-e$):
\begin{equation}
\begin{split}
\ket{\psi}=\frac{1}{2}\left(e^{-i e\int_{\gamma_{A_1}} A^{\mu}dx_{\mu}}b^{\dagger}_{A}+e^{-i e\int_{\gamma_{B_1}} A^{\mu}dx_{\mu}}b^{\dagger}_{B}\right)
\otimes\left(e^{i e\int_{\gamma_{A_2}} A^{\mu}dx_{\mu}}c^{\dagger}_{A}+e^{i e\int_{\gamma_{B_2}} A^{\mu}dx_{\mu}}e^{i\beta}c^{\dagger}_{B}\right)\ket{0},
\end{split}
\end{equation}
where $b^{\dagger}$ and $c^{\dagger}$ are respectively positron and electron creation operators.\\
Next, Alice and Bob let the particles interact and they postselect exclusively on processes which give rise to photons, thereby discarding the one-particle sector and those processes in which the pairs scatter without annihilating (e.g. the Bhabba scattering). The annihilation processes give rise to photon pairs of different momenta $\vec{k}$ and $\vec{k'}$:
\begin{equation}
b^{\dagger}_{i}c^{\dagger}_{i} \rightarrow a^{\dagger}_{i,\vec{k}}a^{\dagger}_{i,\vec{k'}},
\end{equation}
where $a^{\dagger}_{i,\vec{k}}$ denotes a (suitably smeared) creation operator for a photon of momentum $\vec{k}$ produced at location $i$. The quantum state after the interaction and post-selection is thus
\begin{equation}\label{totst}
\ket{\psi}=\frac{1}{\sqrt{2}}\left( a^{\dagger}_{A,\vec{k_A}}a^{\dagger}_{A,\vec{k_A'}}+e^{i\Delta\varphi'}a^{\dagger}_{B,\vec{k_B}}a^{\dagger}_{B,\vec{k_B'}} \right)\ket{0},
\end{equation}
where the phase difference is now
\begin{equation}
\Delta \varphi'=\beta+\varphi'_{B_1}+\varphi'_{B_2}-\varphi'_{A_2}-\varphi'_{A_1}, \quad \varphi_{i_{j}}'\equiv e_j\int_{\gamma_{i_{j}}} A^{\mu}dx_{\mu}.
\end{equation}
Since the electric charges of the two particles are $e_1=-e$ and $e_2=e$, the phase difference acquired due to the interaction with the gauge potential is given by a spacetime loop integral of the gauge potential and is equal to Eq.~\eqref{eq:phase}. Once Alice and Bob possess their photons, they can estimate the phase from measurement results of local projections on states
\begin{equation}
\ket{\pm}_{i}=\frac{1}{\sqrt{2}}\left( \mathbb{1} \pm  a^{\dagger}_{\vec{k_i}}a^{\dagger}_{\vec{k_i'}}  \right)\ket{0}.
\end{equation}
Therefore, we found a different procedure which yields the same gauge invariant phase  as the one obtained in the protocol involving two identical particles.


 \section{Local measurements on a general charged fermionic state}\label{app:C}
Before tackling the fully general case, let us first analyse the following simpler case. Suppose that the scenario involves three sources which emit three electrons in spatial superposition towards three parties labelled as A, B and C. Upon receiving the particles, the parties perform local unitary transformations on their pertaining wave packets, detect the particles and postselect on the cases in which each party detects one excitation. The postselected state of interest (prior to the local transformations) is thus
\begin{equation}
\ket{\psi}_{PS}=\frac{1}{\sqrt{6}} \sum_{\substack{i,j,k=1 \\i\neq j \neq k \neq i}}^{3} e^{i\phi_{ijk}} c^{\dagger}_{A_i}c^{\dagger}_{B_j}c^{\dagger}_{C_k} \ket{0},
\end{equation} 
where $c^{\dagger}$ are fermionic creation operators (for example $c^{\dagger}_{A_i}$ creates the $i$-th particle at A's location). The phase $\phi_{ijk}$ arises due to the coupling to the gauge potential $A^{\mu}$ (for simplicity we omit the mechanical phases):
\begin{equation}
\phi_{ijk}=e\left(\int_{\gamma_{A_i}} A^{\mu}dx_{\mu} + \int_{\gamma_{B_j}} A^{\mu}dx_{\mu} + \int_{\gamma_{C_k}} A^{\mu}dx_{\mu}  \right),
\end{equation}
where e.g. $\gamma_{A_i}$ is the trajectory traced by the first particle towards A's location. The parties apply local unitary transformations 
\begin{equation}
c^{\dagger}_{P_i} \rightarrow \sum_{j=1}^{3} U^{(P)}_{ij}c^{\dagger}_{P_j},\quad P=A,B,C,
\end{equation}
where $U^{(P)}_{ij}$ are matrix elements of the transformations.\\
The final state before the measurement is thus 
\begin{equation}
\ket{\psi}_{PS}=\frac{1}{\sqrt{6}} \sum_{lmn} \sum_{\substack{i,j,k=1 \\i\neq j \neq k \neq i}}^{3} e^{i\phi_{ijk}} U^{(A)}_{il}U^{(B)}_{jm}U^{(C)}_{kn} c^{\dagger}_{A_l}c^{\dagger}_{B_m}c^{\dagger}_{C_n} \ket{0}.
\end{equation}
Finally, the parties perform local projective measurements; the probability of measuring the state $c^{\dagger}_{A_l}c^{\dagger}_{B_m}c^{\dagger}_{C_n}\ket{0}$ is 
\begin{equation}\label{example2}
P(A_lB_mC_n)=\frac{1}{6}|\sum_{\substack{i,j,k=1 \\i\neq j \neq k \neq i}}^{3} e^{i\phi_{ijk}} U^{(A)}_{il}U^{(B)}_{jm}U^{(C)}_{kn}|^2.
\end{equation}
One of the phase differences that appear in Eq. \eqref{example2} is for instance $\phi_{123}-\phi_{312}$, which can be expressed as a sum of two bipartite loop integrals, as shown in Fig. \ref{fig:4}. The analogous holds for all other outcome probabilities, i.e. all information that one can gather from local measurements can be reconstructed from bipartite loop integrals.\\

\begin{figure}[H]
\includegraphics[width=\linewidth]{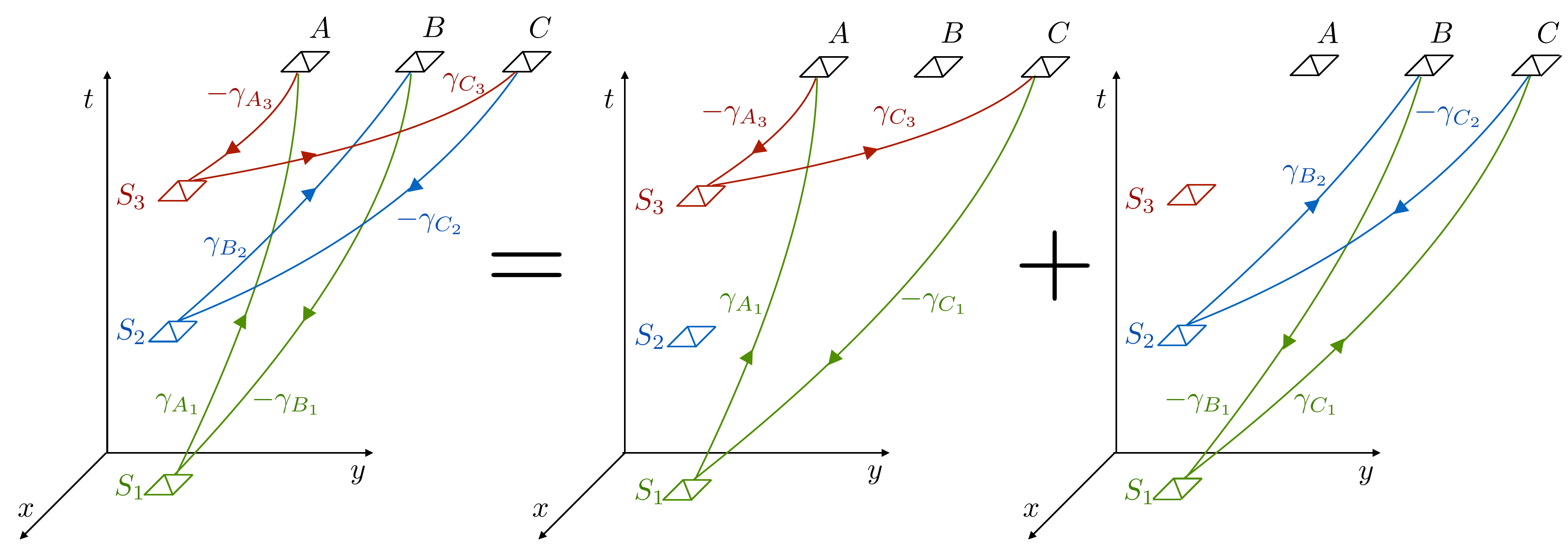}
\caption{The loop integral on the left hand side corresponds to the quantity $\phi_{123}-\phi_{312}$ which can be estimated from one of the outcome probabilities arising in the experiment involving three sources and three parties. The right hand side shows that the latter integral can be decomposed as a sum of two bipartite integrals similar to the ones from Fig. \ref{fig:spacetime_AB}.}
\centering
\label{fig:4}
\end{figure}

Let us now turn to the general case and prove that an analogous decomposition to the one in Fig. \ref{fig:4} can be made for experiments involving any number of sources and any number of parties. Suppose that $d$ sources emit single identical charged fermions (with charge $e$) in arbitrary superpositions of spatial trajectories towards $N$ parties. Adopting the same notation as in Appendix \ref{app:A}, the state shared by the parties upon receiving the particles is thus 
\begin{equation}\label{inital state}
\ket{\psi_0}=\sum_{\substack{\vec{x}_1,...,\vec{x}_N \\ \sum_n x_{nj}=1, \forall j }} \lambda_x e^{i\sum_{n,j}x_{nj}\varphi(n,j)} \prod_{k=1}^{N} \left[ \prod_{j=1}^{d} \left(c_{j}^{(k)\dagger}\right)^{x_{kj}} \right]\ket{0}, \quad \quad \varphi(n,j)\equiv e \int_{\gamma_{n_j}}A^{\mu}dx_{\mu}.
\end{equation}
The bit strings $\left\{\vec{x}_1,...,\vec{x}_N\right\}$ automatically implement the fact that each state can be occupied by at most one fermion; $c^{\dagger}$ are fermionic creation operators (i.e. $c_{j}^{(k)\dagger}$ creates one fermion in the $j$-th mode of $k$-th party), and $\gamma_{n_j}$ indicates the path connecting the $j$-th source to the $n$-th party. Each phase $\varphi(n,j)$ corresponds to the phase picked up by the particle travelling from the $j$-th source towards the $n$-th party. As before, $\lambda_x$ denote normalized real amplitudes.\\
Next, the parties perform local linear operations:
\begin{equation}
c_{i}^{(k)\dagger} \rightarrow \sum_j T_{ij}^{(k)}c_{j}^{(k)\dagger},
\end{equation}
where $T_{ij}^{(k)}$ are arbitrary coefficients (e.g. elements of a unitary matrix). 
The transformed state is thus
\begin{equation}\label{inital state}
\ket{\psi_f}=\sum_{\substack{\vec{x}_1,...,\vec{x}_N \\ \sum_n x_{nj}=1, \forall j }} \lambda_x e^{i\sum_{n,j}x_{nj}\varphi(n,j)} \prod_{k=1}^{N} \left[ \prod_{i=1}^{d} \left(\sum_j T_{ij}^{(k)}c_{j}^{(k)\dagger}\right)^{x_{kj}} \right]\ket{0}.
\end{equation}
Finally, the parties perform local projective measurements giving rise to the following probability distribution
\begin{equation}
P(y)=|\bra{\psi_f} \prod_{k=1}^{N}\left[\prod_{j=1}^{d} \left(c_{j}^{(k)\dagger}\right)^{y_{kj}} \right]\ket{0}|^2.
\end{equation}
After some inspection, one sees that the probabilities involve only interference terms of components with same local particle numbers, i.e. we can write them as
\begin{equation}
P(y)= \sum_{x,x'} \tilde{\lambda}_x\tilde{\lambda}_{x'}^* e^{i\sum_{n,j}(x_{nj}-x_{nj}')\varphi(n,j)}, \quad \quad  \sum_j x_{nj}=\sum_j x_{nj}'=\sum_j y_{nj},\quad \forall n,
\end{equation}
where $\tilde{\lambda}_x$ are coefficients, the exact form of which we leave unspecified. The probabilities thus depend on the following phases
\begin{equation}
\Delta \varphi (x,x')=\sum_{n,j}(x_{nj}-x_{nj}')\varphi(n,j), \quad \sum_{j}(x_{nj}-x_{nj}')=0.
\end{equation}
Now we want to show that any such phase can be written as a combination of bipartite loop integrals like the one shown in Fig. \ref{fig:spacetime_AB}. More precisely, we want to prove that the phases can be cast in the following form
\begin{equation}\label{form}
\sum_{n,n',j,j'} \left[ \varphi(n,j)-\varphi(n,j')+\varphi(n',j')-\varphi(n',j)  \right],
\end{equation}
where each square bracket in the sum represents one such bipartite loop integral.\\
Let us start by introducing the functions $n_j$, such that $x_{kj}=\delta_{k,n_j}$ (this can be done because each source emits one particle which can then be found at most at one party's location). Then, there exists a permutation $\pi$, such that the phase can be written in the following concise form
\begin{equation}
\label{eq:delta_phi}
\Delta \varphi (x,x')=\sum_{j} \left[ \varphi(n_j,j) - \varphi(n_j,\pi(j)) \right].
\end{equation}
The possibility of writing the phase in this form is essentially enabled by the property that states corresponding to $x$ and $x'$ have same local particle numbers: therefore, for every term $(n_j,j)$ arising from bit-string $x$ there must exist a corresponding term $(n_j,f(j))$ arising from bit-string $x'$, where $f$ is a function that maps sources into sources; furthermore, since the particle emitted from each source can be found at only one location, the function $f$ must be one-to-one, and hence a permutation. Now we will prove by construction that the phase can be cast in the form given by \eqref{form}.

Let $G = \langle \pi \rangle$ be the cyclic subgroup of the full permutation group that is generated by $\pi$, where we assume $\pi$ is not the identity permutation. $G$ has a natural group action on the set $\mathbb{Z}_d$. It is well-known that the orbits of a set under the action of a group form a partition of the set; which we denote by $Y := \mathbb{Z}_d / G$. For any equivalence class $[y]$ in $Y$ choose a representative $y$ in $\mathbb{Z}_j$, with the further requirement that the representative of $[1]$ is $1$. Furthermore let $k_y$ be the cardinality of the equivalence class $[y]$; equivalently $k_y$ is the smallest positive integer such that $\pi^{k_y } (y) = y$. With all this notation established, it is straightforward to show that
\begin{equation}
\label{eq:delta_phi_2loop}
\Delta \varphi(x, x') = \sum_{{[y]} \in Y} \sum_{l = 0}^{k_y - 2} \left[ \varphi(n_{\pi^{l}(y)}, y)-\varphi(n_{\pi^{l}(y)},\pi^{l+1}(y))+\varphi(n_{\pi^{l+1}(y)},\pi^{l+1}(y))-\varphi(n_{\pi^{l+1}(y)}, y)  \right].
\end{equation}
The proof proceeds by writing out each term in the sum over $Y$ explicitly and comparing with Eq.~\eqref{eq:delta_phi}. For example the portion of the sum with $y = 1$ is
\begin{equation}
\begin{split}
&\left[ \varphi(n_1, 1)-\varphi(n_1,\pi(1))+\varphi(n_{\pi(1)},\pi(1))-\varphi(n_{\pi(1)},1) \right]\\
&+ \left[\varphi(n_{\pi(1)}, 1)-\varphi(n_{\pi(1)}, \pi^2(1))+\varphi(n_{\pi^2(1)},\pi^2(1))-\varphi(n_{\pi^2(1)},1)\right] + . . . \\
&+ \left[ \varphi(n_{\pi^{k_1-2}(1)}, 1)-\varphi(n_{\pi^{k_1-2}(1)},\pi^{k_1-1}(1))+\varphi(n_{\pi^{k_1-1}(1)},\pi^{k_1-1}(1))-\varphi(n_{\pi^{k_1-1}(1)},1) \right].
\end{split}
\end{equation}
We see that in the sum above, the last term of every bracket is cancelled by the first term of the succeeding bracket. Looking at the third term of every bracket, and the first term of the first bracket, we see that every term $\varphi(n_s, s)$ appears. Finally, the second term of every bracket plus the last term of the last bracket account for all terms $-\varphi(n_s, \pi(s))$.

Thus we have established Eq.~\eqref{eq:delta_phi_2loop}, which shows that $\Delta \varphi (x,x')$, and therefore all probabilities arising from local linear operations/measurements on single charged particles in superposition of spatial trajectories can be deduced from bipartite loop integrals as the one depicted in Fig. \ref{fig:spacetime_AB}.

\end{widetext}

%
%
%

\end{document}